\documentstyle[11pt,a4,epsfig]{article} 
\title{ Impact of Light Higgs Properties on the \\
   Determination of $\tan\beta$
   and $m_{susy}$ 
}

\author{\small
   Jun-ichi {\sc Kamoshita}\footnote{
         E-mail : kamosita@theory.kek.jp}
}

\date{\small
   Department of Physics of Ochanomizu University, Tokyo 112-0012, Japan
}

\begin{document}
\baselineskip 14pt
\maketitle
\begin{abstract}{
     We examine whether parameters related to the Higgs sector of the 
      minimal supersymmetric standard model can be determined by
      detailed study of the production cross section and decay branching
      ratios of the Higgs boson. 
      Assuming that only the light Higgs boson will be observed at a 
      future $e^+e^-$ linear collider with $\sqrt{s}=300\sim500$ GeV,
      we show that values of $m_{susy}$ and $\tan\beta$ are restricted within
      a narrow region in the $m_{susy}$ versus $\tan\beta$ plane by the
      combined analysis of the light Higgs properties.
      It is also pointed out that, in some case, 
      $\tan\beta$ may be restricted
      to a relatively small value, $\tan\beta=1\sim5$.
}\end{abstract}

 The minimal supersymmetric standard model (MSSM) is considered
 to be an attractive candidate as a theory beyond the standard model (SM).
 In the MSSM, the Higgs sector consists of two Higgs doublets, and  
 there exist five physical states: 
 two CP-even Higgs bosons ($h$ and $H$ with $m_h<m_H$),
 one CP-odd Higgs boson $(A)$, and 
 one pair of charged Higgs bosons $(H^\pm)$.
 It is possible to derive specific 
 predictions for this Higgs sector because
 the form of the Higgs potential in the MSSM is very restricted
 in comparison with that in the general two Higgs doublets model.
 In particular, the upper bound on the mass of the lightest CP-even neutral
 Higgs boson is predicted as about 130 GeV.\cite{OYY}
 As for the detectability of the Higgs boson,
 it has been shown that at least one
 CP-even neutral Higgs boson should be detectable at a future $e^+e^-$
 linear collider with $\sqrt{s}=300\sim500$ GeV.\cite{Janot}
 Furthermore, the detectability of the Higgs boson 
 is claimed for a large class of 
 SUSY standard models with extended Higgs sectors.\cite{Kamo}

 Once the Higgs boson is discovered,
 one of the questions of interest is to what extent the parameters
 related to the Higgs sector will be constrained from 
 the detailed study of properties of the Higgs boson.
 By branching ratios of the Higgs boson,
 the mass of a CP-odd Higgs boson ($m_A$) can be constrained 
 almost independently of the SUSY breaking mass scale ($m_{susy}$)
 even if the CP-odd Higgs boson is not discovered 
 at future linear colliders with $\sqrt{s}=300$ GeV.\cite{KOT}

 In this paper we consider the determination of
 parameters of the Higgs sector in the MSSM assuming that 
 only the lightest CP-even Higgs boson will be observed
 at a future $e^+e^-$ linear collider with $\sqrt{s}=300\sim500$ GeV.
 It is shown that the allowed $m_{susy}$-$\tan\beta$ parameter space
 can be restricted within a narrow region by 
 precise measurements of Higgs boson properties.

 Let us begin by listing the parameters of the Higgs sector and 
 the observables which can be used to determine these parameters.
 At the tree level, the masses of Higgs bosons and the mixing angle
 among Higgs bosons are determined by two parameters, 
 the CP-odd Higgs boson mass and the ratio of 
 the vacuum expectation values 
 ($\tan\beta=\frac{\langle H_2 \rangle}{\langle H_1 \rangle}$), where 
  $H_1$ is a Higgs doublet that couples to up-type quarks and 
  $H_2$ is a Higgs doublet that couples to down-type quarks 
                                           and leptons.
 However, once the radiative corrections to the 
 Higgs potential are taken into account, 
 they bring out new parameters in our analysis.
 In the calculation of the Higgs effective potential 
 at the one loop level,
 the most important contribution comes from the top and stop loop, and
 therefore the relevant parameters are 
 two stop masses $(m_{\tilde{t_1}}, m_{\tilde{t_2}})$,
 a Higgsino mass parameter $(\mu)$, and 
 a trilinear soft-breaking parameter $(A_t)$.
 For the moment, we assume that no significant effect is 
 induced from the left-right mixing of
 the stop sector.\footnote{ In our analysis, 
               we use the Higgs mass matrix including the $L$-$R$ mixing effect
               of two stops.\cite{ERZ}}
 Then, effectively 
 there are three parameters related to the Higgs sector.
 As usual,  for these three parameters, 
 we take $m_A$, $\tan\beta$, and  $m_{susy}$ defined by
 $m_{susy}=\sqrt{m_{\tilde{t_1}}m_{\tilde{t_2}}}$.
 Then the CP-even Higgs mass matrix\cite{ERZ} is 
\begin{eqnarray}
   M^2_{\rm higgs}&=&
       \left(\begin{array}{cc} 
         m^2_Z\cos^2\beta+m^2_A\sin^2\beta 
                             & -(m^2_Z+m^2_A)\cos\beta\sin\beta\cr
        -(m^2_Z+m^2_A)\cos\beta\sin\beta &
                  m^2_Z\sin^2\beta+ m^2_A\cos^2\beta 
                                + \frac{\delta_t}{\sin^2\beta}
                        \end{array}\right) ,
\end{eqnarray}
where 
\begin{eqnarray}
   \delta_t=\frac{3m_t^4}{4\pi^2v^2}
                  \ln\left(\frac{m^2_{susy}}{m^2_t}\right) 
\end{eqnarray}
 represents the leading part of the radiative corrections 
 due to the top-stop loop effect, with
 $ v=\sqrt{\langle H_2 \rangle^2+\langle H_1 \rangle^2}\simeq174$ GeV.
 The masses of CP-even Higgs bosons and
 the Higgs mixing angle, $\alpha$, are given by
\begin{eqnarray} 
   m^2_h&=& \frac{1}{2}\left[ 
                  m^2_A+m^2_Z+\delta_t/\sin^2\beta
                      \right. \nonumber \\
        & & \left. 
         -\sqrt{\{(m^2_Z-m^2_A)\cos2\beta-\delta_t/\sin^2\beta\}^2
                    +(m^2_Z+m^2_A)^2\sin^22\beta }
            \right]\label{eqn:mh} ,\\
    m^2_H&=& m^2_A+m^2_Z-m^2_h+\delta_t/\sin^2\beta ,\\
 \tan\alpha&=&\frac{(m^2_Z+m^2_A)\cos\beta\sin\beta}
                 {m^2_h-(m^2_Z\cos^2\beta+m^2_A\sin^2\beta)} \ .
\end{eqnarray}

 The lightest CP-even Higgs boson is mainly produced through
 the Higgs-strahlung process, $e^+e^-\to Zh$, 
 at a $e^+e^-$ linear collider with $\sqrt{s}=300\sim500$ GeV.
 If we assume that the decay modes of the Higgs boson to SUSY particles
 are not dominant,\footnote{ 
             If decays of the Higgs boson to 
          the SUSY particles are observed, 
          we can see obviously that 
          the Higgs boson belongs to the SUSY model.
          We will not consider such a case because
          we are now interested in the case that 
          the SM-like Higgs boson will be observed.
           }
 then the main decay mode of the Higgs boson is 
 the $h\to b\bar{b}$ mode.
 In this case, 
 the behavior of the Higgs boson may be similar to 
 that of a Higgs boson in the SM.
 The lightest Higgs boson then has sizable decay branching ratios
 in the modes $h\to b\bar{b}, \tau\bar{\tau}, c\bar{c}$
 and $gg$.\footnote{Since the availability of $h\to WW^{\ast}$ depends 
                    crucially on the Higgs boson mass, we will not 
                    consider this mode here.}

 With a reasonable luminosity of $\sim50$ fb$^{-1}$/year, 
 the mass of the Higgs boson, $m_h$, can be determined precisely
 by the recoil mass distribution.\cite{JLC,WCHB,NLCHWG}
 The Higgs production cross section,
 $\sigma(e^+e^-\to Zh)$, is obtained by
 the branching ratio of the $Z$ boson decaying into $l\bar{l}(l=e, \mu)$ and 
 the cross section of the event with 
 the recoil mass around $m_h$.\cite{WCHB}
 The production cross section multiplied by 
 the branching ratio of $h\to X (X=\{b\bar{b}\}, \{\tau\bar{\tau}\}, 
 \{c\bar{c}$~{\scriptsize or}~$gg\})$,\footnote{Although 
               it is very difficult to measure the branching ratios of
               the modes $c\bar{c}$ and $gg$ separately,
               the sum of $Br(h\to c\bar{c})$ and $Br(h\to gg)$ can be
               measured with reasonable
               precision.\cite{WCHB,NLCHWG,HBB,Nakamura}
               We denote the sum of $Br(h\to c\bar{c})$ and $Br(h\to gg)$ as 
               $Br(h\to c\bar{c}\ { }_{\rm or}\ gg)$.}
 $\sigma(e^+e^-\to Zh)Br(h\to X)$,
 can be obtained by the $ZX$ production rate with 
 the invariant mass of $X$ being around $m_h$.\cite{WCHB,NLCHWG}

\begin{figure}[t]
 \begin{center}
   \epsfxsize = 5 cm
   \centerline{\epsfbox{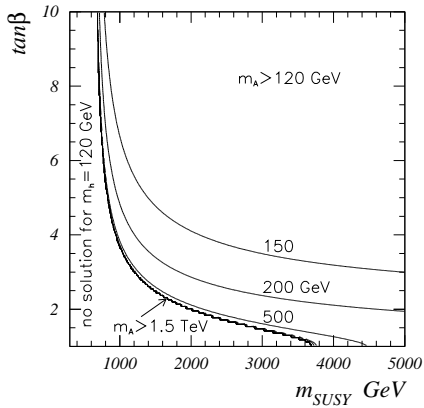}}
  \begin{minipage}{14.5cm}
   \vspace{-10mm}
   \caption[]{ Contour plots of the value of $m_A$ for $m_h=120$ GeV are 
        shown in the $m_{susy}$ versus $\tan\beta$ plane.
        We take the top quark mass as $m_t=175$ GeV.
        The Higgs mass formula Eq.(\ref{eqn:mh}) can not be satisfied 
        for $m_h=120$ GeV in
        the left and bottom left region in the figure. 
        In large $\tan\beta$ and large $m_{susy}$ region,
        the value of $m_A$ is always larger than 120 GeV 
        when $m_h=120$ GeV.  }   \label{fign:ma}
  \end{minipage}
 \end{center}
\end{figure}

 The three parameters 
 $m_A,\tan\beta$ and $m_{susy}$
 will be restricted by the observables mentioned above.
 Expected experimental errors of observables 
 have been estimated in detail.\cite{NLCHWG}
 According to there estimates, 
 the error of $m_h$ should be 0.1 $\sim$ 0.5\%. 
 Therefore, in the following,
 we treat the value of $m_h$ as fixed.
 Thus there are two remaining degrees of freedom of parameters.
 Hereafter we choose $m_{susy}$ and $\tan\beta$ as free parameters 
 and derive the value of $m_A$ with 
 the Higgs mass formula Eq.~(\ref{eqn:mh}) for the fixed value of $m_h$.
 For $m_h=120$ GeV, Fig.~\ref{fign:ma} displays the contour plot of $m_A$ 
 in the $m_{susy}$-$\tan\beta$ plane.

\begin{table}[t]
\begin{center}
\begin{minipage}{14.5cm}
\caption[]{
         Couplings of the light Higgs boson to a fermion pair in 
         the MSSM and the SM.
         $u,d$ and $l$ represents up-type quarks ($u=\{u,c,t\}$),
                                down-type quarks ($d=\{d,s,b\}$), 
                             and leptons ($l=\{e,\mu,\tau\}).$ 
            }\label{tb:cpl}
\end{minipage}
\begin{tabular}{ c|c|c|c }
\hline
\hline
        & $h$-$u$-$u$ & $h$-$d$-$d$ & $h$-$l$-$l$ \\
\hline
        & & & \\

 MSSM   & $-i\frac{\mbox{$m_u$}}{\mbox{$v$}}
             \frac{\displaystyle\cos\alpha}{\displaystyle\sin\beta}$ 
        & $ i\frac{\mbox{$m_d$}}{\mbox{$v$}}
             \frac{\displaystyle\sin\alpha}{\displaystyle\cos\beta}$ 
        & $ i\frac{\mbox{$m_l$}}{\mbox{$v$}}
             \frac{\displaystyle\sin\alpha}{\displaystyle\cos\beta}$ \\
        & & & \\
  SM    & $-i\frac{\mbox{$m_u$}}{\mbox{$v$}}$ 
        & $-i\frac{\mbox{$m_d$}}{\mbox{$v$}}$ 
        & $-i\frac{\mbox{$m_l$}}{\mbox{$v$}}$ \\
\end{tabular}
\end{center}
\end{table}

 The ratio of branching ratios,
 for example
 $Br(h\to~c\bar{c}$~{\scriptsize or}~$gg)/Br(h\to~b\bar{b})$,
 will be determined with reasonable precision.\cite{WCHB,NLCHWG,Nakamura}
 The formulas for the partial decay width of 
 the Higgs boson in the MSSM are derived, for example,
 in Ref.~\cite{Barger}. 
 Higgs-fermion-fermion couplings are listed in 
 Table~\ref{tb:cpl}.
 The partial decay width for 
 $h\to b\bar{b}$ and $h\to\tau\bar{\tau}$
 are proportional to the down-type fermion-Higgs coupling, and then
 the ratio $Br(h\to~\tau\bar{\tau})/Br(h\to~b\bar{b})$ is 
 the same as that in the SM. Therefore no information on 
 the parameters of the Higgs sector in the MSSM are obtained from 
 this ratio.\footnote{
                    This ratio is important to determine the bottom mass,
                    as discussed, for example, in Refs.~\cite{KOT,NLCHWG}.
                    }
 On the other hand, as reported in Ref.~\cite{KOT}, the ratio  
 $Br(h\to c\bar{c}$~{\scriptsize or}~$gg)/Br(h\to b\bar{b})$
 is a useful variable to constrain the value of $m_A$,
 because the ratio 
 strongly depends on $m_A$ but is almost independent of $m_{susy}$.

 The determination of $\tan\beta$ 
 has great implication for both the theoretical and 
 experimental study of SUSY standard models,
 because not only the physics of the Higgs sector 
 but also that of other SUSY sectors, 
 for example the chargino and neutralino sector, depend on $\tan\beta$. 
 Therefore we must start to use other observables
 in order to determine the values of both $m_{susy}$ and $\tan\beta$.

\begin{figure}[t]
 \begin{center}
 \begin{tabular}{lcr}
  \begin{minipage}{4.5cm}
   \epsfxsize = 4.5 cm
   \centerline{\epsfbox{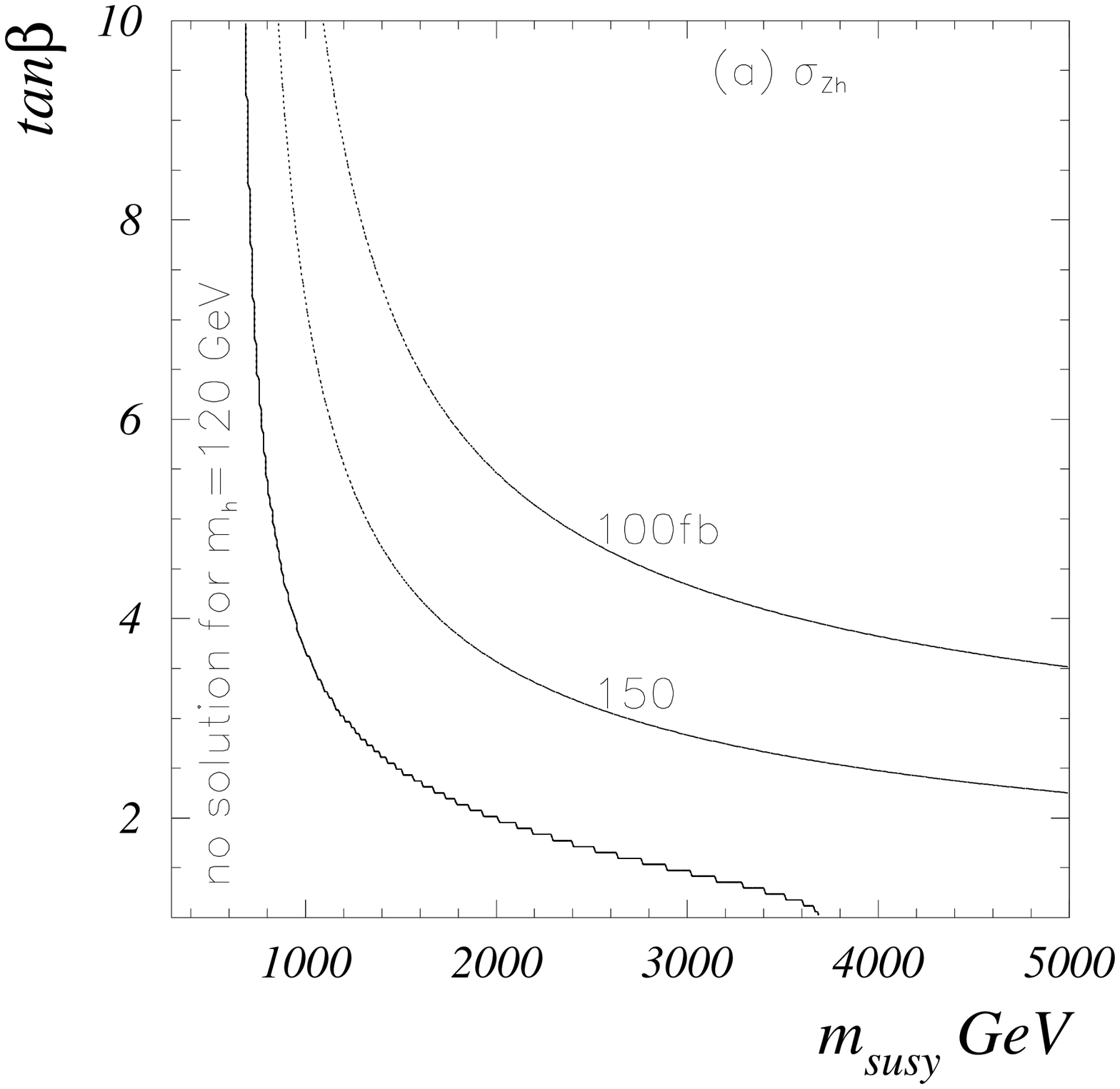}}
  \end{minipage} 
&
  \begin{minipage}{4.5cm}
   \epsfxsize = 4.5 cm
   \centerline{\epsfbox{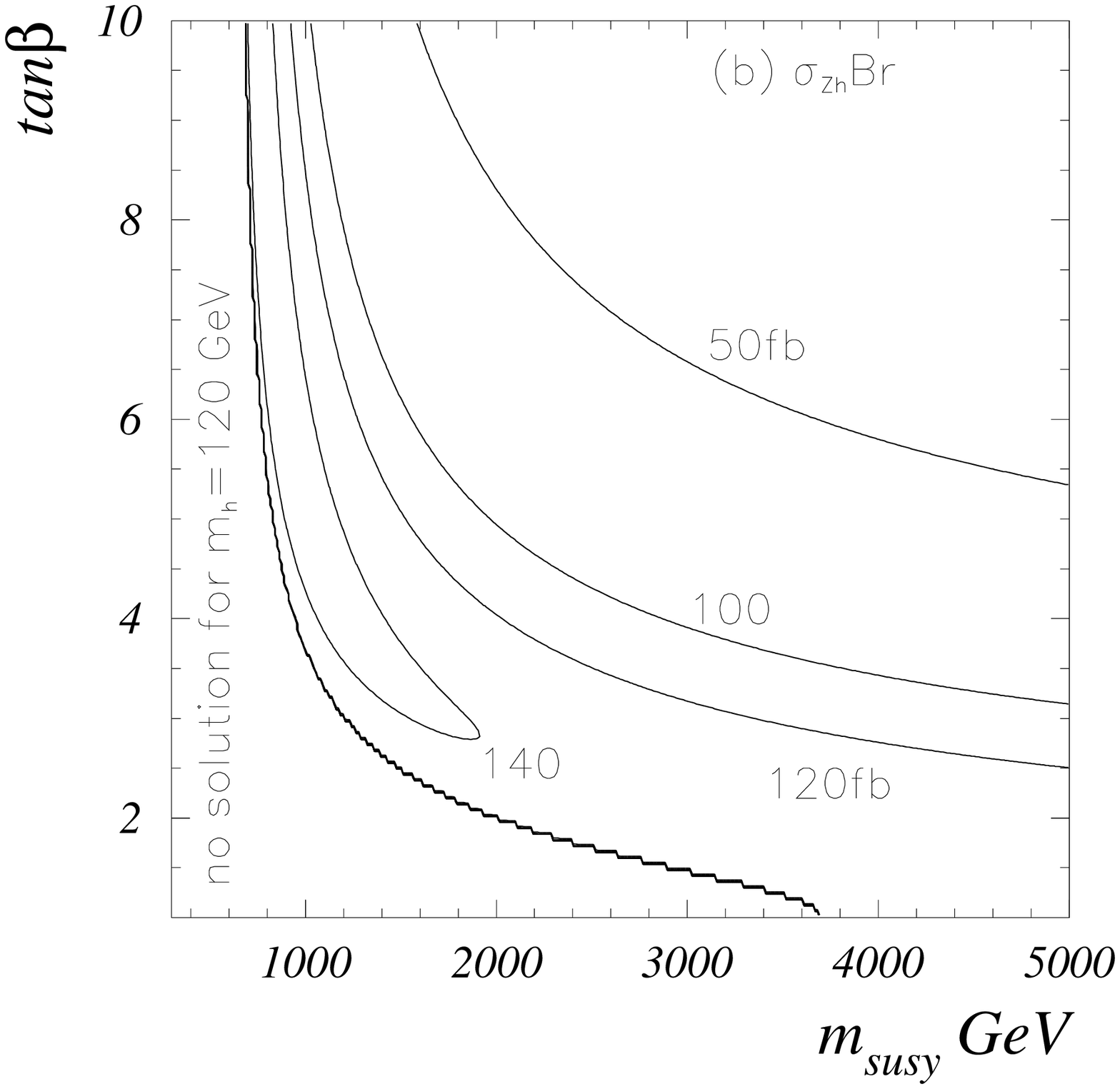}}
  \end{minipage} 
&
  \begin{minipage}{4.5cm}
   \epsfxsize = 4.5 cm
   \centerline{\epsfbox{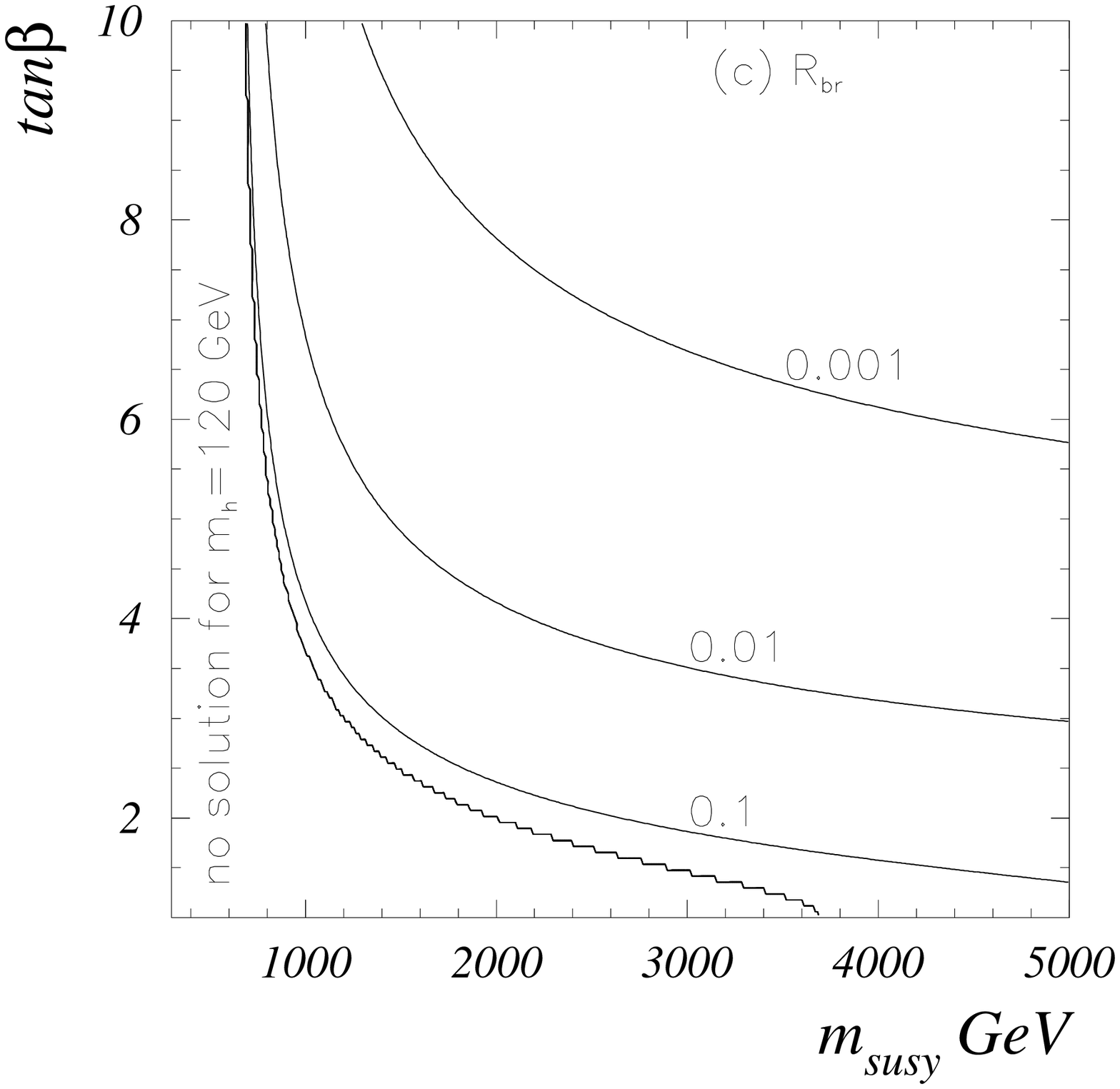}}
  \end{minipage} 
 \end{tabular}
\begin{minipage}{14.5cm}
   \vspace{-10mm}
   \caption[]{Contours plots of (a) $\sigma_{Zh}$, 
        (b) $\sigma_{Zh}Br(b\bar{b})$, and (c) $R_{br}$ are shown.
        We take the quark masses as 
        $m_t=175$ GeV, $m_b(m_b)=4.2$ GeV and $m_c(m_c)=1.3$ GeV.
        The strong coupling constant is taken as
        $\alpha_s(m_Z)=0.12$.  }   \label{fign:obs}
\end{minipage}
 \end{center}
\end{figure}

 Hereafter we use abbreviated notation defined as follows:
 $\sigma_{Zh}\equiv\sigma(e^+e^-\to Zh)$,
 $\sigma_{Zh}Br(b\bar{b})\equiv\sigma(e^+e^-\to Zh)Br(h\to b\bar{b})$
 and 
  $R_{br}\equiv
  Br(h\to c\bar{c}$~{\scriptsize or}~$gg)/Br(h\to b\bar{b})$.
 These observables give us 
 different constraints on the values of $m_{susy}$ and $\tan\beta$.
 As discussed in Ref.~\cite{KOT}, 
 we obtain the approximate relation
 \begin{eqnarray}
  R_{br} \propto\left(\frac{1}{\tan\beta\tan\alpha}\right)^2
  \nonumber. 
 \end{eqnarray}
 Both $\sigma(e^+e^-\to Zh)$ and $\sigma(e^+e^-\to Zh)Br(h\to~b\bar{b})$ 
 depend on the angles $\alpha$ and $\beta$ as
 \begin{eqnarray}
  \sigma_{Zh}  &\propto&\sin^2(\alpha-\beta) , \nonumber\\
  \sigma_{Zh}Br(b\bar{b}) &\propto&  
                \sin^2(\alpha-\beta)\left(
                    1 + \frac{m_\tau^2}{3m_b^2} + R_{br} + f(\alpha,\beta)
                                    \right)^{-1}
    \nonumber ,
 \end{eqnarray}
 where 
 \begin{eqnarray}
 f(\alpha,\beta)\equiv
     \left(\Gamma_{\rm tot}-\sum_X\Gamma(h\to X)\right)/\Gamma(h\to b\bar{b}),
 \end{eqnarray}
 $X=b\bar{b},\tau\bar{\tau},c\bar{c},gg$.
 Here $\Gamma_{\rm tot}$ is the total decay width of the light Higgs boson.
 In Fig.~\ref{fign:obs}(a)$\sim$(c) display the contour plots of 
 $\sigma_{Zh}$, $\sigma_{Zh}Br(b\bar{b})$ and $R_{br}$, respectively,
 in the $m_{susy}$-$\tan\beta$ plane for $m_h=120$GeV.
 The shape of the contours in Fig.~\ref{fign:obs}(b)
 is somewhat different from the other two
 in the left side of the figure.
 Fig.~\ref{fign:obs}(a) for $\sigma_{Zh}$ is similar to
 Fig.~\ref{fign:obs}(c) for $R_{br}$.
 However, Fig.~\ref{fign:obs}(a) displays gentle slope,
 as compared with Fig.~\ref{fign:obs}(c).

 Now we combine these observables to estimate the constraints on 
 the values of $m_{susy}$ and $\tan\beta$.
 For this purpose, 
 we take $m_{susy}$ and $\tan\beta$ as fitting parameters 
 and then  perform the $\chi^2$ test in 
 the $m_{susy}$-$\tan\beta$ plane 
 for a fixed value of $m_h$.
 The value of $m_A$ is derived from 
 the Higgs mass formula Eq.(\ref{eqn:mh})
 point-by-point in the $m_{susy}$-$\tan\beta$ plane.

 However, we input the values of $m_h$, $m^0_A$ and $m^0_{susy}$ 
 as true values for the $\chi^2$
 test,\footnote{
       In order to distinguish ``true" values of  $m_A$ and $m_{susy}$
       from $m_A$ and $m_{susy}$ as fitting parameters, 
       the index ``$0$" is appended to the ``true" values.
                }
 because these variables have clear physical meanings as 
 the mass of particles and a typical mass scale for $m_{susy}$.
 As for $\tan\beta$, the ``true" value is calculated from
 the input parameters, $m_h$, $m^0_A$ and $m^0_{susy}$, 
 by the Higgs mass formula Eq.~(\ref{eqn:mh}).

\begin{table}[t]
\begin{center}
\begin{minipage}{14.5cm}
\caption[]{
     List of errors for each observable discussed in Ref.~\cite{NLCHWG}. 
     $R_{br}$ is defined by
     $R_{br}\equiv Br(h\to c\bar{c}$~{\scriptsize or}~$gg)/Br(h\to b\bar{b})$
\cite{KOT}.
            }\label{tb:error}
\end{minipage}
\begin{tabular}{c|c|c|c|c} 
\hline
\hline
  $m_h$   & $\delta(m_h)$ & $\delta(\sigma_{Zh})$ 
                  & $\delta(\sigma_{Zh}Br(b\bar{b}))$ 
                  & $\delta(R_{br})$  \\
\hline
  110 GeV & $0.1\sim 0.5$\% & $\sim 7$\% & $\sim 2.5$\% &$\sim 14$\% \\
  120 GeV & $0.1\sim 0.5$\% & $\sim 7$\% & $\sim 3.5$\% &$\sim 14$\% \\
\hline
\vspace{3mm}
\end{tabular}
\end{center}
\end{table}

 Definition of $\chi^2$ is given by
 \begin{eqnarray}
  \chi^2\equiv&&
      \left\{\left(\frac{\sigma_{Zh}-\sigma_{Zh}^{0}}
                               {\delta\sigma_{Zh}}\right)^2 
      +\left(\frac{\sigma_{Zh}Br(b\bar{b})
                     -\sigma_{Zh}Br(b\bar{b})^{0}}
                         {\delta(\sigma_{Zh}Br(b\bar{b}))}\right)^2 
      +\left(\frac{R_{br}-R_{br}^{0}}{\delta(R_{br})}\right)^2
             \right\}, 
  \label{eqn:chisq}
 \end{eqnarray} 
 where $\delta(\sigma_{Zh})$, $\delta(\sigma_{Zh}Br(b\bar{b}))$ and 
 $\delta(R_{br})$ represent expected experimental errors.
 The estimated error of each observable reported in Ref.~\cite{NLCHWG}
 is summarized in Table \ref{tb:error}.
 $\sigma_{Zh}^{0}$, $\sigma_{Zh}Br(b\bar{b})^{0}$ and $R_{br}^{0}$ are 
 the central values derived from the input parameters, 
 $m_h, m_A^0$ and $m_{susy}^0$.
 The values of $\sigma_{Zh}$, $\sigma_{Zh}Br(b\bar{b})$ and $R_{br}$ are 
 calculated at each point in the  
 $m_{susy}$ versus $\tan\beta$ plane.
 To calculate the Higgs production cross section,
 we use $\sqrt{s}=350$GeV.\footnote{
              Of course when $m_A<\sqrt{s}/2$,
        the CP-odd Higgs boson will be produced by an
        associated production process, $e^+e^-\to AH$.
        In this case we can use many observables depending on 
        SUSY parameters and 
        should convert the strategy of our analysis to another one. 
        Since we assume that only a light Higgs boson will be discovered,
        we constrain our analysis to $m_A>\sqrt{s}/2\sim 180$ GeV.
        Hereafter we will not consider the case $m_A<180$ GeV. 
              }

\begin{figure}[t]
 \begin{center}
  \begin{tabular}{cc}
   \begin{minipage}{60mm} \qquad (a)
   \end{minipage} 
 & 
   \begin{minipage}{60mm} \qquad \qquad (b)
   \end{minipage} 
\cr
   \begin{minipage}{60mm} \vspace{-2cm}
   \epsfxsize = 6 cm
   \centerline{\epsfbox{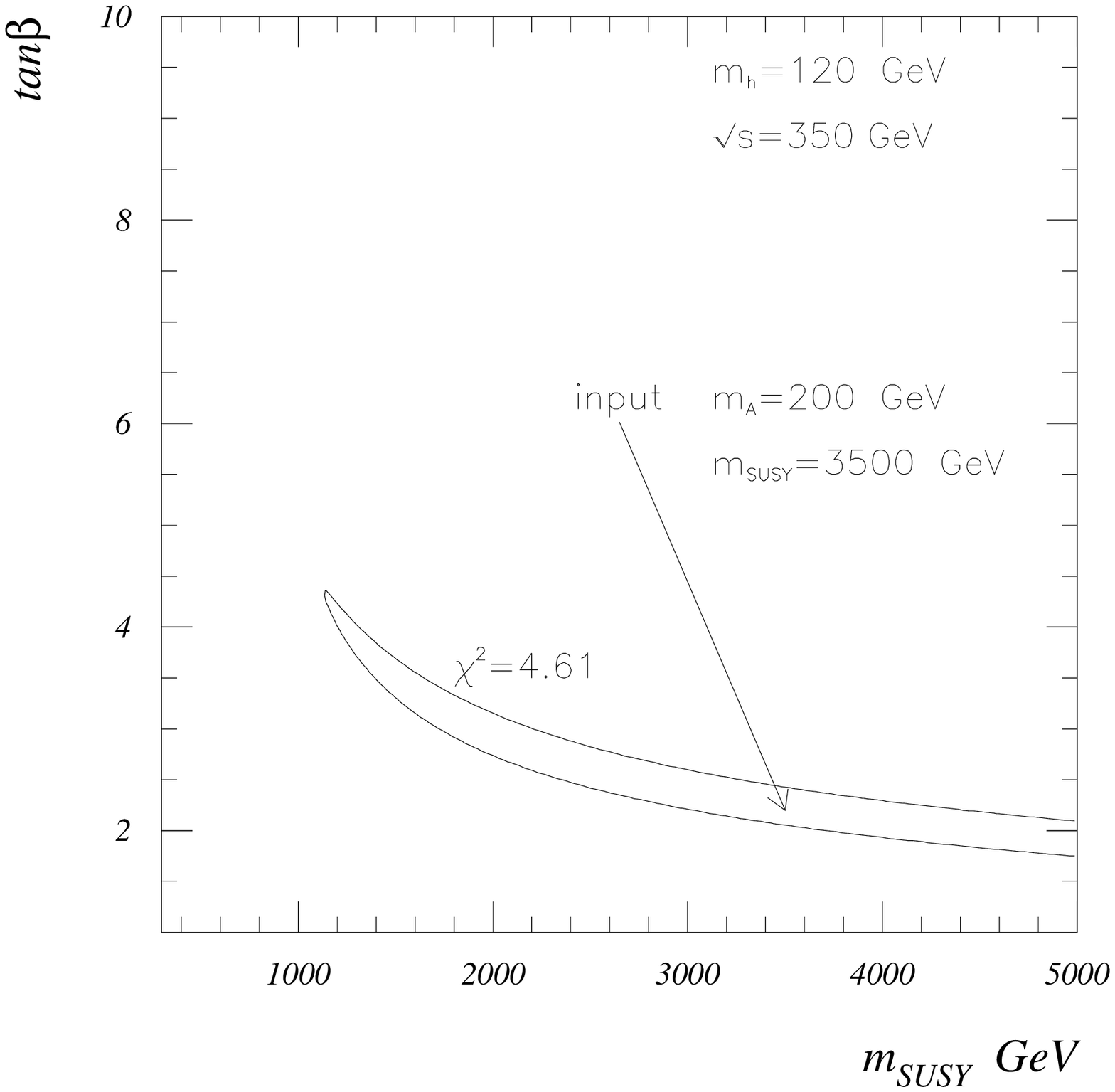}}
   \end{minipage} 
 &
   \begin{minipage}{60mm}\vspace{-2cm}
   \epsfxsize = 6 cm
   \centerline{\epsfbox{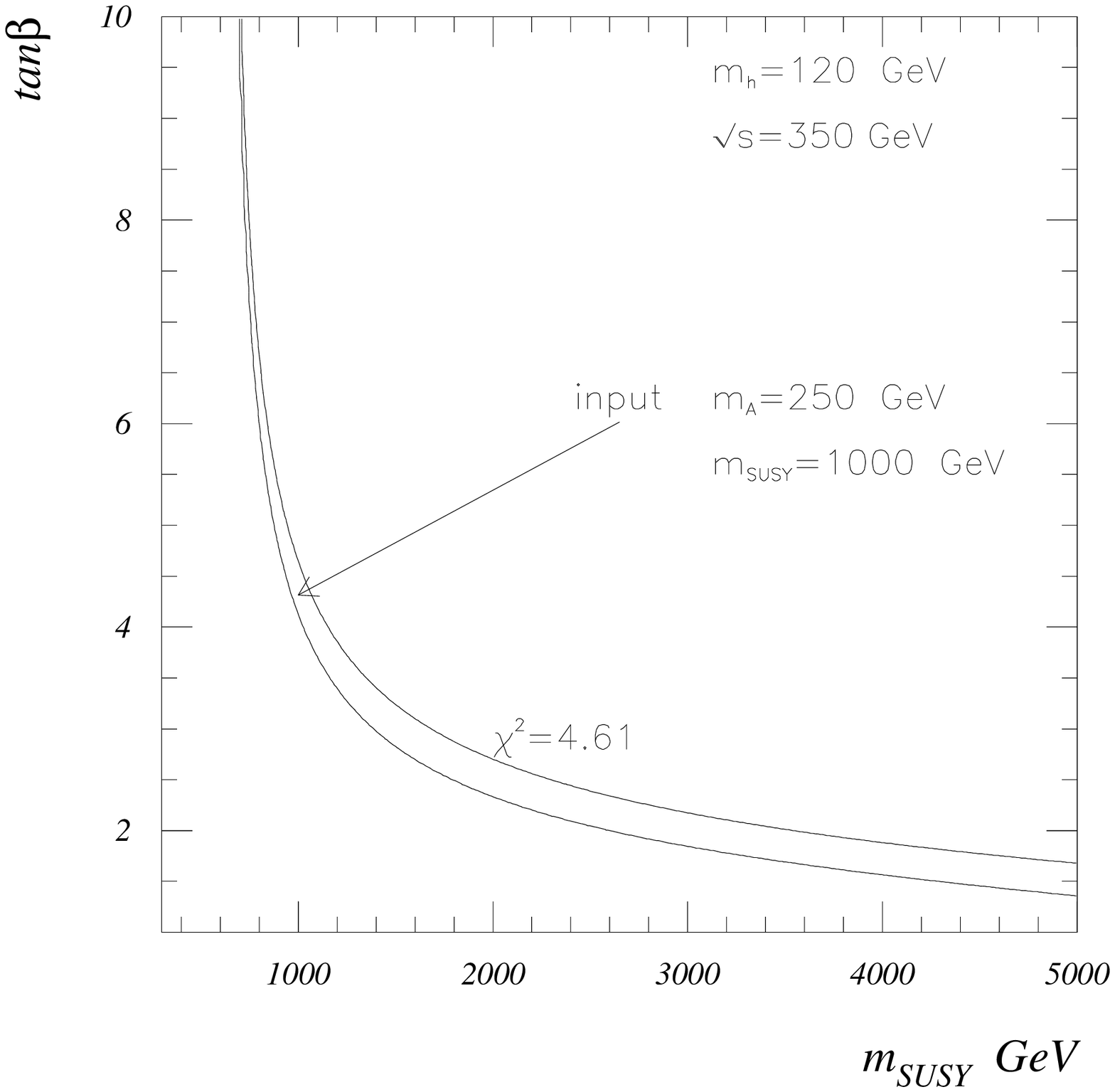}}
   \end{minipage} 
  \end{tabular}
\begin{minipage}{14.5cm}
  \vspace{-10mm}
   \caption[]{ Contour plot of $\chi^2$ with $\chi^2=4.61$
        (a) for 
        ($m_h, m^0_A, m^0_{susy}$)=(120 GeV, 200 GeV, 3500 GeV) and 
        (b) for 
        ($m_h, m^0_A, m^0_{susy}$)=(120 GeV, 250 GeV, 1000 GeV).
         $\chi^2<4.61$ inside a narrow region.}  \label{fign:ch2i}
\end{minipage}
 \end{center}
\end{figure}

 The contour plots of $\chi^2$ for $m_h=120$ GeV are shown in 
 Figs.~\ref{fign:ch2i}(a) and (b) with a 95\%CL contour. 
 We find in Fig.~\ref{fign:ch2i}(a) that the $\tan\beta$ is restricted 
 within a relatively small value, $\tan\beta<4.5$, and 
 the value of $m_{susy}$ is weakly restricted, $m_{susy}>1$ TeV.
 Fig.~\ref{fign:ch2i}(b) displays the contour plot of $\chi^2$ 
 for other input value.
 In Fig.~\ref{fign:ch2i}(b), 
 although  the upper bounds on $m_{susy}$ and $\tan\beta$ 
 are not obtained in the displayed region,
 the allowed $m_{susy}$-$\tan\beta$ parameter space is restricted within
 a narrow region.

\begin{figure}[t]
 \begin{center}  
   \begin{tabular}{lll}
   \begin{minipage}{4.5cm}
   \epsfxsize = 4.5 cm
   \centerline{\epsfbox{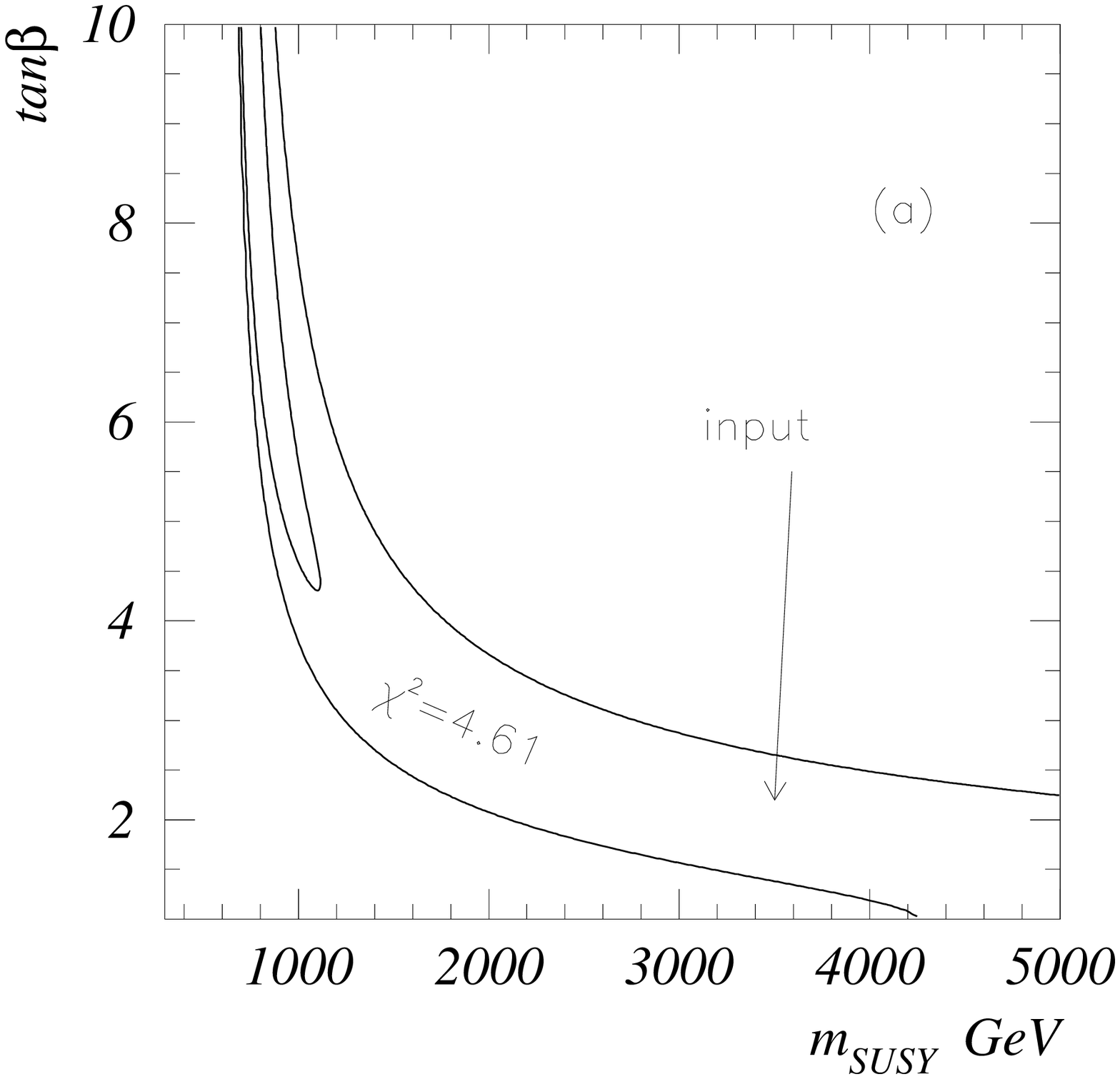}}
   \end{minipage} &
   \begin{minipage}{4.5cm}
   \epsfxsize = 4.5 cm
   \centerline{\epsfbox{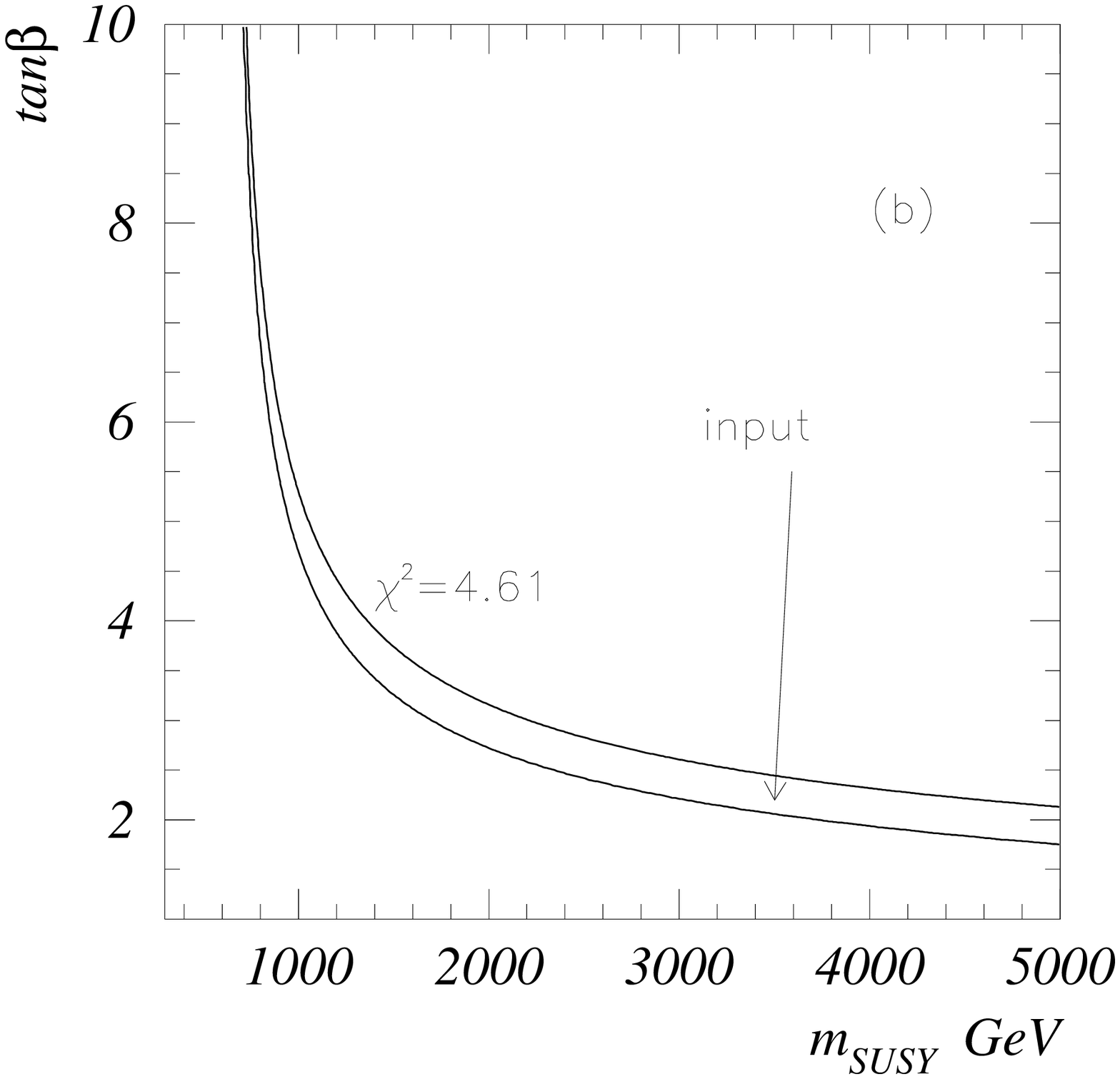}}
   \end{minipage} &
   \begin{minipage}{4.5cm}
   \epsfxsize = 4.5 cm
   \centerline{\epsfbox{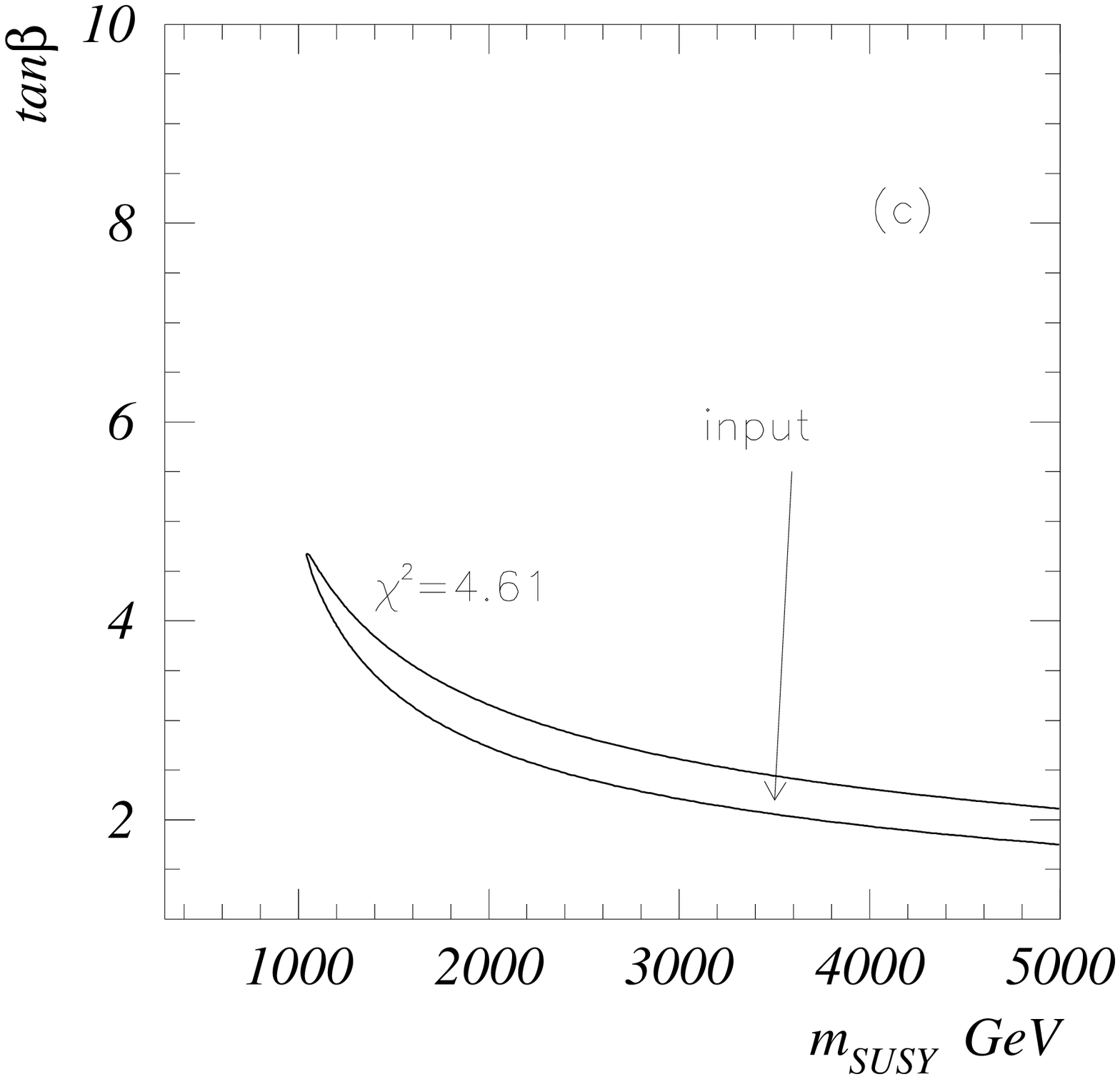}}
   \end{minipage}
   \end{tabular}
\begin{minipage}{14.5cm}
   \vspace{-10mm}
   \caption[]{ Contour plots of $\chi^2$ with $\chi^2=4.61$
        when we use just two observables among 
        $\sigma_{Zh}$, $\sigma_{Zh}Br(b\bar{b})$ and $R_{br}$:
        (a) $\sigma_{Zh}$ and $\sigma_{Zh}Br(b\bar{b})$,
        (b) $\sigma_{Zh}$ and $R_{br}$,
        (c) $\sigma_{Zh}Br(b\bar{b})$ and $R_{br}$.
        Input values of the parameters are taken to be
        the same as in Fig. 3(a). }   \label{fign:twof3}
\end{minipage}
  \end{center}
\end{figure}

 Next, in order to show how each observable contributes to constrain
 the $m_{susy}$-$\tan{\beta}$ parameter space,
 we show the $\chi^2$ contour plots in Fig.~\ref{fign:twof3}
 by using just two observables among the three observables.
 We can see from Fig.~\ref{fign:twof3} that $R_{br}$
 contributes strongly to the constraint on 
 the $m_{susy}$-$\tan\beta$ plane.

\begin{figure}[th]
  \begin{center}
   \epsfxsize = 6 cm
   \centerline{\epsfbox{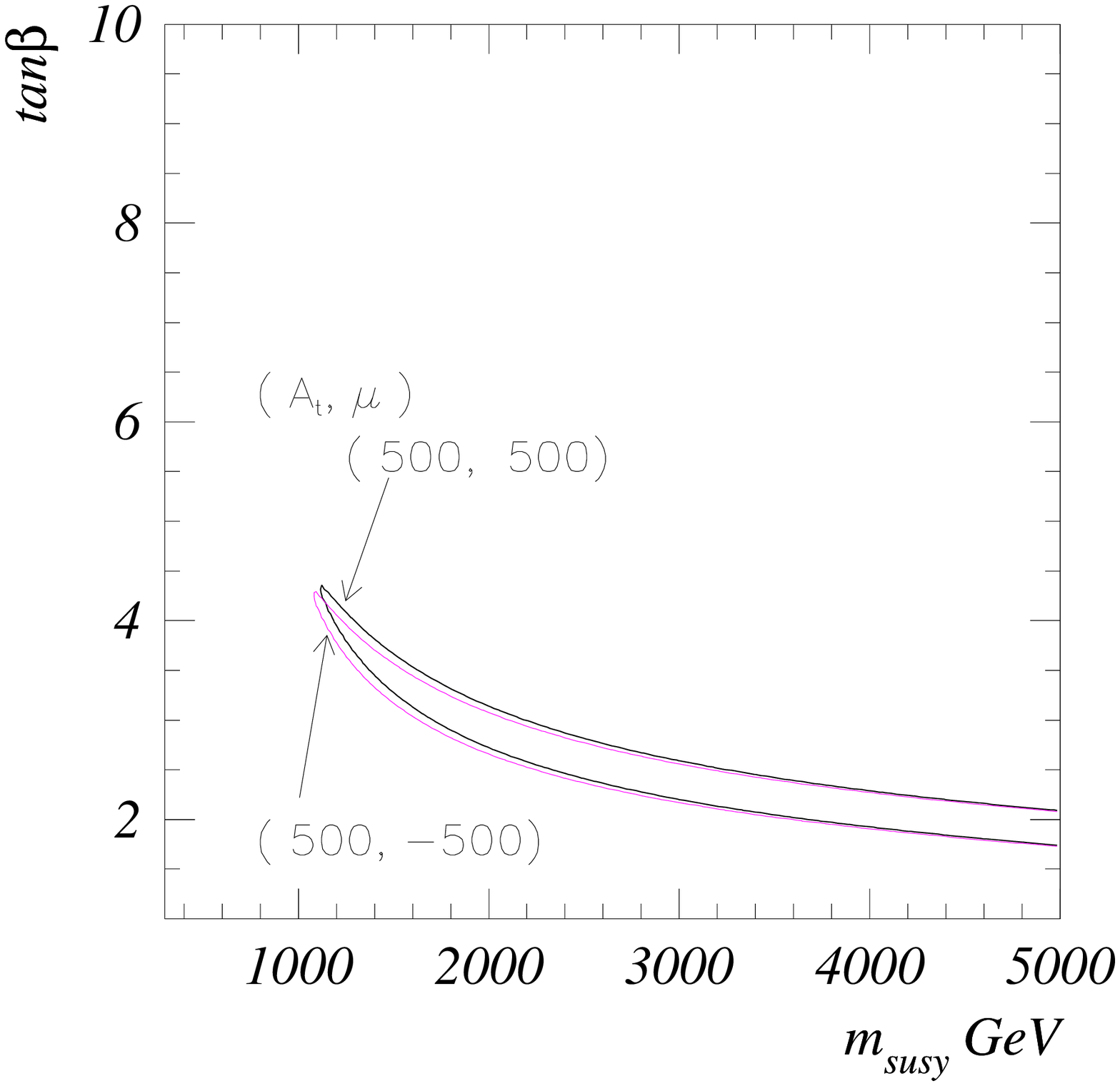}}
\begin{minipage}{14.5cm}
   \vspace{-10mm}
   \caption[]{ Contour plots of $\chi^2$ with $\chi^2=4.61$
        including the L-R mixing effect of two stops.
        The values shown in the parenthesis represent ($A_t$, $\mu$) in GeV.
        Input values of other parameters are taken to be
        the same as in Fig. 3(a). }   \label{fign:LR}
\end{minipage}
\end{center}
\end{figure}

 The results above can be understood as follows.
 Once the value of $m_h$ is fixed, $\tan\beta$ and $m_{susy}$ are 
 strongly correlated by the Higgs mass formula Eq.~(\ref{eqn:mh}).
 We can consider Fig.~\ref{fign:ma} as showing 
 the value of $\tan\beta$ as a function of $m_{susy}$
 for fixed values of $m_h$ and $m_A$. From Fig.~\ref{fign:ma},
 the $m_{susy}$-$\tan\beta$ parameter space is restricted within
 a relatively narrow region 
 even if $m_A$ varies from $\sim 200$ GeV to larger than 1 TeV.
 However the constraint obtained from Fig.~\ref{fign:ma}
 is weak as compared with that from Fig.~\ref{fign:ch2i}(a) and (b).
 We can see from Fig.~\ref{fign:twof3} that $R_{br}$
 contributes strongly to the constraint on 
 the $m_{susy}$-$\tan\beta$ plane.
 In Figs.~\ref{fign:ch2i}(a)~and~\ref{fign:twof3},
 the value of $m_A$ is restricted within 
 about 180$ \sim $230 GeV by $R_{br}$, and as a result
 the region satisfying the constraints becomes narrow
 as compared with that obtained in Fig.~\ref{fign:ma}.
 The reason why the upper bound on $\tan\beta$ is obtained in 
 Fig.~\ref{fign:ch2i}(a) is, in addition to $R_{br}$, 
 $\sigma_{Zh}Br(b\bar{b})$ contributes effectively to 
 the constraint on the $m_{susy}$-$\tan\beta$ plane. 


  So far, we have neglected the $L$-$R$ mixing of the stop sector. 
  We can include the $L$-$R$ mixing effects and
  take non-zero values of $A_t$ and $\mu$
  in our analysis.
  In this case, examples are shown in Fig.~\ref{fign:LR}.
  The contour should be shifted by
  varying the values of $A_t$ and $\mu$.
  However, the result of our analysis will not 
  change essentially, because
  $R_{br}$ is almost independent of 
  the parameters of the stop sector, as shown in Ref.~\cite{KOT}. 

 With regard to theoretical aspects, 
 the requirement of Yukawa coupling unification in 
 SUSY-GUT\cite{YKWU} restricts the value of $\tan\beta$
 to two solutions.
 One is the small $\tan\beta$ solution, $\tan\beta=1\sim 3$, 
 and the other one is the large $\tan\beta$ solution, $\tan\beta\sim 50$.
 There are mainly two types of scenarios for 
 Yukawa coupling unification.
 These are the bottom-tau Yukawa unification and 
 the top-bottom-tau Yukawa unification scenarios.
 The requirement of bottom-tau Yukawa coupling unification suggests
 both the small $\tan\beta$ solution and the large $\tan\beta$ solution. 
 However, the requirement of top-bottom-tau Yukawa coupling 
 unification suggests only the large $\tan\beta$ solution.
 Therefore, if a large value of $\tan\beta$ will be excluded by 
 precise measurements of light Higgs properties at a 
 future linear collider, for example
 as we have shown in Fig.~\ref{fign:ch2i}(a),
 the experiments may rule out
 the top-bottom-tau Yukawa unification scenario
 even if only the lightest CP-even Higgs boson is observed.
 
 We now conclude our discussion.
 We have examined 
 whether the parameters of the Higgs sector in the MSSM
 can be determined by detailed study of Higgs properties.
 We have found that 
 the values of $\tan\beta$ and $m_{susy}$ are restricted
 within a very narrow region even if
 only the light Higgs boson is discovered. 
 $R_{br}$ contributes strongly to the constraint on
 the $m_{susy}$-$\tan\beta$ plane.
 We also have shown that
 the upper bound on $\tan\beta$ may be obtained  
 by combining analysis of observables
 when $\sigma_{Zh}Br(b\bar{b})$ contributes effectively to the constraint.
 However, to obtain a more strict constraint on 
 both $m_{susy}$ and $\tan\beta$, 
 we need constraints inferred from other quantities
 obtained from heavy Higgs and/or SUSY particles.  

\vspace{2mm}
 The author would like to thank A. Sugamoto, S. Kamei and M. Aoki for
 reading the manuscript and helpful comments.

\end{document}